\documentclass[aip,prd,showpacs,nofootinbib]{revtex4}
\usepackage{latexsym}
\usepackage{amsfonts}
\usepackage{amsbsy}
\usepackage{mathrsfs}

\begin{document}

\title{BF gravity with Immirzi parameter and cosmological constant}

\date{\today}

\author{Merced Montesinos}\email{merced@fis.cinvestav.mx}

\author{Mercedes Vel\'azquez}\email{mquesada@fis.cinvestav.mx}

\affiliation{Departamento de F\'{\i}sica, Cinvestav, Instituto Polit\'ecnico
Nacional 2508, San Pedro Zacatenco, 07360, Gustavo A. Madero, Ciudad de
M\'exico, M\'exico.}

\begin{abstract}
The action principle of the BF type introduced by Capovilla, Montesinos,
Prieto, and Rojas (CMPR) which describes general relativity with Immirzi
parameter is modified in order to allow the inclusion of the cosmological
constant. The resulting action principle is on the same footing as the
original Plebanski action in the sense that the equations of motion coming
from the new action principle are equivalent to the Holst action principle
plus a cosmological constant without the need of imposing additional
restrictions on the fields. We consider this result a relevant step towards
the coupling of matter fields to gravity in the framework of the CMPR action
principle.
\end{abstract}


\maketitle

The research in quantum gravity led by its two main branches (loop quantum
gravity \cite{lqg} and spin foam models for gravity \cite{spinfoam}) has
recently motivated the study of the classical descriptions for general
relativity and theories related to it, particularly, the formulations of
gravity of the BF type. For instance, Cartan's equations in the framework of
BF theories are analyzed in Refs. \cite{vlad1,vlad2} while the relationship of
general relativity to the Husain-Kuchar model in the framework of BF theory is
analyzed in Refs. \cite{meche1,meche2,cuando}.

As is well know, general relativity was formulated as a constrained BF theory
by Pleba\'nski in the mid-seventies \cite{jmp1977}. The fundamental variables
used for the description of the gravitational field are two-form fields, a
connection one-form, and some Lagrange multipliers. This framework was
extended much later in order to include the coupling of matter fields
\cite{cdjm}. These formulations are complex and one has to use reality
conditions in order to get real general relativity. There are other
formulations for gravity expressed as a constrained BF theory which are real,
and were given in Refs. \onlinecite{mike,cqg1999b} and \onlinecite{cqgl2001}.
In particular, the one given in Ref. \onlinecite{cqgl2001} includes the
so-called Immirzi parameter \cite{barbero,holst,immirzi}. Recently, the issue
of the introduction of the cosmological constant in the framework of Ref.
\onlinecite{cqgl2001} was studied by Smolin and Speziale \cite{simone}. This
issue is also the goal of this paper. Our approach is different from the one
followed in Ref. \onlinecite{simone} and, in our opinion, is close to the
original Pleba\'nski formulation. So, we are ready to make the introduction of
the cosmological constant and, at the end of the paper, we will comment on the
relationship between the results of this paper and the ones of Ref.
\onlinecite{simone}.

The action principle for pure gravity introduced by Capovilla, Montesinos,
Prieto, and Rojas in Ref. \onlinecite{cqgl2001} (hereafter CMPR) is given by
\begin{eqnarray}\label{cmpr}
S [Q,A,\psi,\mu] =\int_{\mathscr{M}^4} \left [ Q^{IJ} \wedge F_{IJ} [A] -
\frac12 \psi_{IJKL} Q^{IJ} \wedge Q^{KL} - \mu  \left ( a_1\, \psi_{IJ}\,^{IJ}
 + a_2\, \psi_{IJKL}\, \varepsilon^{IJKL} \right ) \right ],
\end{eqnarray}
where $A^I\,_J$ is a Euclidean or Lorentz connection one-form, depending on
whether $SO(4)$ or $SO(3,1)$ is taken as the internal gauge group, and
$F^I\,_J[A]= d A^I\,_J + A^I\,_K \wedge A^K\,_J$ is its curvature; the $Q$'s
are a set of six two-forms on account of their antisymmetry $Q^{IJ}=-Q^{JI}$;
the Lagrange multiplier $\psi_{IJKL}$ has 21 independent components due to the
properties $\psi_{IJKL}=\psi_{KLIJ}$, $\psi_{IJKL}=-\psi_{JIKL}$, and
$\psi_{IJKL}=-\psi_{IJLK}$; the Lagrange multiplier $\mu$ implies the
additional restriction $a_1\, \psi_{IJ}\,^{IJ} + a_2\, \psi_{IJKL}\,
\varepsilon^{IJKL}=0$ on the Lagrange multiplier $\psi_{IJKL}$. The Lorentz
(Euclidean) indices $I,J,K\ldots$ are raised and lowered with the Minkowski
(Euclidean) metric $(\eta_{IJ})= \mbox{diag}(\sigma,+1,+1,+1)$, where
$\sigma=+1$ for Euclidean and $\sigma=-1$ for Lorentzian signatures.

The variation of the action (\ref{cmpr}) with respect to the independent
fields gives the equations of motion
\begin{eqnarray}
&& \delta Q: F_{IJ} [A] - \psi_{IJKL} Q^{KL}=0, \label{delta Q}\\
&& \delta A: D Q^{IJ} =0, \label{delta w}\\
&& \delta \psi: Q^{IJ} \wedge Q^{KL} + 2 a_1 \mu\, \eta^{[I\mid K \mid}
\eta^{J]L} + 2 a_2 \mu\, \varepsilon^{IJKL}=0,\label{delta psi}\\
&& \delta \mu: a_1\, \psi_{IJ}\,^{IJ} + a_2\, \psi_{IJKL} \varepsilon^{IJKL} =0. \label{delta mu}
\end{eqnarray}
It was shown in Ref. \onlinecite{cqgl2001} that the solution of Eq.
(\ref{delta psi}) for the two-forms $Q^{IJ}$ is given by
\begin{eqnarray}\label{sol}
Q^{IJ}=\alpha\,  {^{\ast}} \left ( e^I \wedge e^J \right ) + \beta\, e^I \wedge e^J,
\end{eqnarray}
where ${^{\ast}}\left ( e^I \wedge e^J \right )= \frac12
\varepsilon^{IJ}\,_{KL} e^K \wedge e^L$ and the constants $\alpha$ and $\beta$
satisfy
 \begin{eqnarray}\label{quotien}
\frac{a_2}{a_1} &=& \frac{\alpha^2 + \sigma \beta^2}{4 \alpha\beta}.
\end{eqnarray}
By inserting the solution (\ref{sol}) into (\ref{cmpr}), we get
\begin{eqnarray}\label{cmprEnes}
S[e,A] = \alpha \int_{\mathscr{M}^4} \left [ {^{\ast}} \left ( e^I
\wedge e^J \right ) +  \frac{\beta}{\alpha}\, e^I \wedge e^J  \right
] \wedge F_{IJ} [A].
\end{eqnarray}

Before introducing the cosmological constant, it is important to remark the
logic involved in the BF-like description for gravity given in Eq.
(\ref{cmpr}). First of all, once we know the meaning of $Q$'s through Eq.
(\ref{sol}), the insertion of (\ref{sol}) in (\ref{delta w}) implies that
$A^I\,_J$ is the spin connection. Therefore, the curvature $F^I\,_J$ must
satisfy Bianchi identities with no torsion $F^I\,_J \wedge e^J=0$ or,
equivalently, using $F_{IJ}=\frac12 F_{IJKL} e^K \wedge e^L$,
\begin{eqnarray}\label{bianchi}
F_{IJKL} + F_{IKLJ} + F_{ILJK}=0.
\end{eqnarray}
Einstein's equations in vacuum, on the other hand, are given by
${^{\ast}F}^I\,_J \wedge e^J=0$ or, equivalently,
\begin{eqnarray}\label{einstein}
{^{\ast}F}_{IJKL} + {^{\ast}F}_{IKLJ} + {^{\ast}F}_{ILJK}=0,
\end{eqnarray}
where ${^{\ast}F}_{IJKL}= \frac12 \varepsilon_{IJ}\,^{MN} F_{MNKL}$ is the
left dual of $F_{IJKL}$. Therefore, in vacuum, Einstein's equations
(\ref{einstein}) have the same form as the Bianchi identities (\ref{bianchi})
but with $F_{IJKL}$ replaced by its left dual ${^{\ast}F}_{IJKL}$. The
relevant point in the BF description given in (\ref{cmpr}) is that the Eqs.
(\ref{bianchi}) impose additional restrictions on the field $\psi_{IJKL}$ and
once these are taken into account Eqs. (\ref{einstein}) are automatically
satisfied due to the use of Eq. (\ref{delta Q}), the properties of
$\psi_{IJKL}$, and Eq. (\ref{delta mu}).

Now, let us come back to the issue of the cosmological constant, the heart of
this paper. The inclusion of the cosmological constant $\Lambda$ modifies the
right-hand side of (\ref{einstein}), which becomes
\begin{eqnarray}\label{einstein+cc}
{^{\ast}F}_{IJKL} + {^{\ast}F}_{IKLJ} + {^{\ast}F}_{ILJK}= \Lambda \varepsilon_{IJKL},
\end{eqnarray}
while (\ref{bianchi}) remains if we assume that $F^I\,_J$ is still the
curvature of the spin connection, i.e., that there is no torsion.  In order to
introduce $\Lambda$ in the CMPR action principle we have at our hands the two
volume terms that can be built with the $Q$'s: $Q_{IJ} \wedge Q^{IJ}$ and
$Q_{IJ} \wedge {^{\ast}Q}^{IJ}$; so it looks natural to use both of them
following Ref. \onlinecite{mon2}. With this in mind, we consider the modified
CMPR action principle
\begin{eqnarray}\label{cmpr mod}
S [Q,A,\psi,\mu] =\int_{\mathscr{M}^4} \left [ Q^{IJ} \wedge F_{IJ} [A] -
\frac12 \psi_{IJKL} Q^{IJ} \wedge Q^{KL} - \mu  \left ( a_1\, \psi_{IJ}\,^{IJ}
+ a_2\, \psi_{IJKL}\, \varepsilon^{IJKL} -\mathcal{H}\right ) \right. \\ \nonumber
\left. + l_1\, Q_{IJ} \wedge Q^{IJ} + l_2\, Q_{IJ} \wedge {^{\ast}Q}^{IJ} \right ],
\end{eqnarray}
where $l_1$, $l_2$, and $\mathcal{H}$ are constants. As it will be apparent
next, the action principle (\ref{cmpr mod}) describes general relativity with
cosmological constant $\Lambda$; the relationship among $l_1$, $l_2$,
$\mathcal{H}$, and $\Lambda$ will be also displayed.

Some of the consequences of the new action principle (\ref{cmpr mod}) is that
the equations of motion obtained from the variation of (\ref{cmpr mod}) with
respect to the independent fields $Q^{IJ}$ and $\mu$ are, instead of Eqs.
(\ref{delta Q}) and (\ref{delta mu}), now given by
\begin{eqnarray}
&& \delta Q: F_{IJ} [A] - \psi_{IJKL} Q^{KL} + 2 l_1\, Q_{IJ} + 2 l_2\, {^{\ast}Q}_{IJ}=0, \label{delta Q mod}\\
&& \delta \mu: a_1\, \psi_{IJ}\,^{IJ} + a_2\, \psi_{IJKL} \varepsilon^{IJKL} -\mathcal{H} = 0, \label{delta mu mod}
\end{eqnarray}
while the variation of (\ref{cmpr mod}) with respect to $A^I\,_J$ and
$\psi_{IJKL}$ gives the same equations as before, namely Eqs. (\ref{delta w})
and (\ref{delta psi}). Therefore, Eqs. (\ref{sol}) and (\ref{quotien}) are
still valid for the modified action (\ref{cmpr mod}), and so by plugging into
(\ref{cmpr mod}) the expression for $Q^{IJ}$ in terms of the tetrad $e^I$, the
action principle becomes
\begin{eqnarray}\label{cmpr mod En es 1}
S[e,A]=\int_{\mathscr{M}^4}\left[ \left[ \alpha\, {^{\ast}}\left( e^I \wedge e^J \right)
 + \beta\, e^I \wedge e^J \right] \wedge F_{IJ} + \mu \mathcal{H}
 + \left( l_1\, \alpha \beta + \frac{l_2}{2}
 \left(\alpha^2 \sigma+ \beta^2\right) \right ) \varepsilon_{IJKL}
 e^I\wedge e^J \wedge e^K \wedge e^L \right].
\end{eqnarray}
Furthermore, by inserting (\ref{sol}) into (\ref{delta Q mod}) the curvature
acquires the form
\begin{eqnarray}\label{efe}
F_{IJKL}= 2\alpha {\psi^{\ast}}_{IJKL} + 2\beta \psi_{IJKL}-2(l_1\alpha +
l_2 \beta) \varepsilon_{IJKL} -2 (l_1 \beta + l_2
\alpha \sigma)(\eta_{IK}\eta_{JL}-\eta_{IL}\eta_{JK}),
\end{eqnarray}
with ${\psi^{\ast}}_{IJKL}= \frac12 \varepsilon_{KL}\,^{MN} \psi_{IJMN}$ being
the right dual of $\psi_{IJKL}$. Since Eqs. (\ref{sol}) and (\ref{delta w})
still hold for the action principle (\ref{cmpr mod}) and they imply that
$A^I\,_J$ is the spin connection, then $F_{IJKL}$ must satisfy the Bianchi
identities (\ref{bianchi}), like in the case without cosmological constant.
Therefore, Eqs. (\ref{bianchi}) and (\ref{efe}) imply the following additional
restrictions,
\begin{eqnarray}\label{Bi epsilon}
2\alpha \left( {\psi^{\ast}}^I\,_{JKL} + {\psi^{\ast}}^I\,_{KLJ}+
{\psi^{\ast}}^I\,_{LJK}\right) +2\beta \left( \psi^I\,_{JKL} +
\psi^I\,_{KLJ}+\psi^I\,_{LJK}\right) -6(l_1\alpha+l_2\beta)\varepsilon^I\,_{JKL}=0,
\end{eqnarray}
on the components of the Lagrange multipliers $\psi_{IJKL}$, which allow us to
rewrite the curvature $F_{IJKL}$ (\ref{efe}) as
\begin{eqnarray}\label{F con bianchi}
F_{IJKL}= 2\alpha {\psi^{\ast}}_{IJKL}+ 2\beta \psi_{IJKL}-
\frac23 \alpha \left( {\psi^{\ast}}_{IJKL} + {\psi^{\ast}}_{IKLJ}+ {\psi^{\ast}}_{ILJK}\right)
- \frac23 \beta \left( \psi_{IJKL} + \psi_{IKLJ}+
\psi_{ILJK}\right)\nonumber \\ -2(l_1 \beta+ l_2 \alpha \sigma)
(\eta_{IK}\eta_{JL}-\eta_{IL}\eta_{JK}).
\end{eqnarray}
By plugging (\ref{F con bianchi}) into (\ref{bianchi}) it is clear that the
Bianchi identities (\ref{bianchi}) are satisfied, as it has to be, on account
of the properties of $\psi_{IJKL}$.

Up to here, the Bianchi identities have been incorporated in the formalism.
The remaining thing to do is to check whether Einstein's equations with
cosmological constant (\ref{einstein+cc}) are indeed satisfied or not. In
fact, the fulfilling of (\ref{einstein+cc}) can be accomplished by properly
relating the constants ${\mathcal H}$, $l_1$, and $l_2$ with $\Lambda$. This
is done as follows. First of all, the contraction of (\ref{Bi epsilon}) with
$\varepsilon_{IJKL}$ implies a linear equation for the two invariants
$\psi_{IJ}\,^{IJ}$ and $\psi_{IJKL} \varepsilon^{IJKL}$, given by
\begin{eqnarray}\label{bi 4}
\alpha \sigma \psi_{IJ}\,^{IJ} +\beta{\psi^{\ast}}_{IJ}\,^{IJ}-12\sigma (l_1\alpha +l_2\beta)=0.
\end{eqnarray}
Nevertheless, we also have the Eq. (\ref{delta mu mod}) that also relates
linearly the two invariants. There is no way to avoid the fulfilling of these
two equations, the two must be satisfied. By combining them and using
(\ref{quotien}) it is possible to obtain
\begin{eqnarray}\label{combina bi4 y delta mu}
\beta \psi_{IJ}\,^{IJ} +\alpha\, {\psi^{\ast}}_{IJ}\,^{IJ} =
2\beta\frac{\mathcal{H}}{a_1} -12 l_1 \beta -12 l_2 \frac{\beta^2}{\alpha},
\end{eqnarray}
which is, by construction, not independent of (\ref{delta mu mod}) and
(\ref{bi 4}). Equation (\ref{combina bi4 y delta mu}) is a key point that
allows us to include the cosmological constant in the CMPR framework because,
on the other hand, Einstein's equations (\ref{einstein+cc}) are automatically
satisfied due to the fulfilling of the Bianchi identities [or, equivalently,
due to the fact that (\ref{Bi epsilon}) or (\ref{F con bianchi}) holds] with
the only exception, given by
\begin{eqnarray}\label{E4}
\beta \psi_{IJ}\,^{IJ} +\alpha\, {\psi^{\ast}}_{IJ}\,^{IJ}= 12
(l_1\beta +l_2\alpha\sigma) + 2\Lambda.
\end{eqnarray}
Nevertheless, by comparing (\ref{combina bi4 y delta mu}) and (\ref{E4}), it
is easy to see that (\ref{E4}) is satisfied by choosing for ${\mathcal H}$ the
value
\begin{eqnarray}\label{H}
\mathcal{H}=a_1 \left[ 4! l_2 \sigma \frac{a_2}{a_1} +12 l_1 +\frac{\Lambda}{\beta} \right],
\end{eqnarray}
that is to say, the choice for ${\mathcal H}$ given in Eq. (\ref{H}) assures
that Einstein's equations (\ref{einstein+cc}) are automatically satisfied.

Finally, notice that when the expressions for the $Q$'s given in Eq.
(\ref{sol}) and for $\mathcal{H}$ given in Eq. (\ref{H}) are plugged into the
action principle (\ref{cmpr mod}), it acquires the form in terms of the tetrad
fields and the Lorentz (Euclidean) connection
\begin{eqnarray}\label{cmpr mod En es final}
S[e,A] = \alpha \int_{\mathscr{M}^4} \left\{ \left [ {^{\ast}} \left ( e^I
\wedge e^J \right ) +  \frac{\beta}{\alpha}\, e^I \wedge e^J  \right
] \wedge F_{IJ} [A] -\frac{\Lambda}{12}\, \varepsilon_{IJKL}
e^I \wedge e^J \wedge e^K \wedge e^L \right\},
\end{eqnarray}
as expected; where it was used in the intermediate steps that $\mu
a_1=-\frac1{12}\,Q^{IJ}\wedge Q_{IJ}$ and $\mu
a_2=-\frac{\sigma}{24}\,Q^{IJ}\wedge Q^{\ast}\,_{IJ}$, which come from
(\ref{delta psi}).

In summary, our results are collected in the following items: (a) We have
shown that by choosing the expression for $\mathcal{H}$ given in Eq. (\ref{H})
and by inserting it into the action principle (\ref{cmpr mod}), the equations
of motion that follow from this action principle are precisely Einstein's
equations with cosmological constant $\Lambda$ exactly in the same way as the
Plebanski action gives the equations of motion for self-dual gravity with a
cosmological constant. (b) There is no need to impose by hand additional
restrictions on the field variables involved to reach item (a) because the
result comes out solely from the handling of the equations of motion. (c) One
important feature of our construction is that we can even set $l_1=0=l_2$ from
the very beginning into the action (\ref{cmpr mod}) (and so ${\mathcal
H}=\frac{a_1}{\beta} \Lambda$) and still get the equations of motion for
general relativity with cosmological constant $\Lambda$. This fact is
essentially allowed by the equation given in (\ref{delta mu mod}), which is a
modification to the equation of the pure gravity case given in Eq. (\ref{delta
mu}). (d) Other particular cases are given by setting $l_1=0$ and $l_2 \neq 0$
or by $l_1 \neq 0$ and $l_2 =0$ with the appropriate modification for ${\cal
H}$ using (\ref{H}).

To conclude, we compare our approach with the one followed in Ref.
\onlinecite{simone}. The main difference between our approach and theirs is
given in Eq. (\ref{delta mu mod}) of this paper. This fact is crucial to get
precisely Einstein's equations with cosmological constant $\Lambda$. The
authors of Ref. \onlinecite{simone} do not use Eq. (\ref{delta mu mod}),
rather, they use Eq. (\ref{delta mu}). This fact is apparently related to the
fact that they get a cosmological constant which is a function of the Immirzi
parameter. This feature is not present in our approach. Furthermore, even
though they use the two volume terms constructed out of the $Q$'s, in our
approach we can get rid of them by considering the particular case in which
$l_1=0$ and $l_2=0$ from the very beginning and still introduce the
cosmological constant.

We think our results might be of interest for the quantum gravity community
and also for the practitioners of general relativity.\\

Warm thanks to Riccardo Capovilla for very fruitful discussions on the subject
of this paper. This work was supported in part by CONACYT, Mexico, Grant No.
56159-F.

\end{document}